\documentclass[12pt]{article}
\usepackage{eurosym}
\usepackage{amssymb}

\input{tcilatex}
\begin{document}

\bigskip \ 

\bigskip \ 

\begin{center}
\textbf{SCALE-FREE GROWING NETWORKS AND GRAVITY}

\smallskip \ 

\textbf{\ }

\textbf{\ }

\smallskip \ 

J. A. Nieto\footnote{%
nieto@uas.uasnet.mx, janieto1@asu.edu}

\smallskip

\textit{Facultad de Ciencias F\'{\i}sico-Matem\'{a}ticas de la Universidad
Aut\'{o}noma de Sinaloa, 80010, Culiac\'{a}n Sinaloa, M\'{e}xico}

\bigskip \ 

\bigskip \ 

\bigskip \ 

Abstract
\end{center}

We propose a possible relation between complex networks and gravity. Our
guide in our proposal is the power-law distribution of the node degree in
network theory and the information approach to gravity. The established
bridge may allow us to carry geometric mathematical structures, which are
considered in gravitational theories, to probabilistic aspects studied in
the framework of complex networks and \textit{vice versa}.

\bigskip \ 

\bigskip \ 

\bigskip \ 

\bigskip

\ 

\bigskip \ 

\bigskip \ 

\bigskip \ 


Pacs numbers: 04.60.-m, 04.65.+e, 11.15.-q, 11.30.Ly

November, 2012

\newpage

\noindent \textbf{1}. \textbf{Introduction}

\bigskip \ 

Random networks with complex topology describe a wide range of systems in
Nature [1]-[2]. Recent advances in this scenario show that most large
networks can be described by mean-field method applied to a system with
scale-free features. In fact, it is found that in the case of scale-free
random networks, the observed power-law degree distribution is%
\begin{equation}
P(k)\sim \frac{1}{k^{\gamma }},  \tag{1}
\end{equation}%
where $P(k)$ is the probability that a vertex in the network is connected to 
$k$ other vertices and $\gamma $ is a numerical parameter called
connectivity distribution exponent. In fact, $\gamma $ is a scale-free
parameter in the sense does not dependent on a characteristic scale of the
network.

Our main goal in this article is to see whether expression (1) can be
related to gravitational arena. If this is the case then we may argue that
we have found a link between complex networks and gravity. Of course, the
idea to see gravity as a some kind of network system is in fact no new,
since goes back to the work of Penrose [3] (see also Refs. [4]-[9]). In this
case the concept of spin networks describes a combinatorial picture of the
geometry of space-time. However most efforts in this direction is
concentrated in the idea to see gravity as spin network. Here, we shall show
that it is not necessary to introduce the spin concept to establish such a
link. We will do this by taking recourse of the connection proposed in Ref.
[10] between gravity and information theory.

\smallskip \ 

\noindent \textbf{2. Complex networks}

\smallskip \ 

Random networks with complex topology [1]-[2] is based in two principles:

\smallskip \ 

(1) \textit{Growth}: starting with small number of vertices $v_{0}$, at
every time step $t$ one adds a new vertex with $e$ (\TEXTsymbol{<}$v_{0}$)
edges (that will be connected to the the vertices already present in the
system).

\smallskip \ 

(2) \textit{Preferential attachment}: When choosing the vertices to which
the new vertex connects, one assumes that the probability $\Pi (k_{i})$ that
a new vertex will be connected to vertex $i$ depends on the connectivity
(node degree) $k_{i}$ of that vertex. Specifically, one has

\begin{equation}
\Pi (k_{i})=\frac{k_{i}}{\sum \limits_{j=1}^{v_{0}+t-1}k_{j}}.  \tag{2}
\end{equation}%
Observe that the sum in (2) goes over all vertices in the system except the
newly introduced one.

Assuming that $k_{i}$ is continuous parameter one can write

\begin{equation}
\frac{\partial k_{i}}{\partial t}=e\Pi (k_{i}).  \tag{3}
\end{equation}%
Thus, considering (2) we have

\begin{equation}
\frac{\partial k_{i}}{\partial t}=\frac{ek_{i}}{\sum%
\limits_{j=1}^{v_{0}+t-1}k_{j}}.  \tag{4}
\end{equation}%
Since

\begin{equation}
\sum \limits_{j=1}^{v_{0}+t-1}k_{j}=2et,  \tag{5}
\end{equation}%
we get formula

\begin{equation}
\frac{\partial k_{i}}{\partial t}=\frac{k_{i}}{2t},  \tag{6}
\end{equation}%
whose solution, with the correct initial condition, is given by

\begin{equation}
k_{i}(t)=e(\frac{t}{t_{i}})^{1/2}.  \tag{7}
\end{equation}

It is important to observe that in general one has

\begin{equation}
\frac{\partial }{\partial t}\sum \limits_{j=1}^{v_{0}+t-1}k_{j}\neq \sum
\limits_{j=1}^{v_{0}+t-1}\frac{\partial k_{j}}{\partial t}.  \tag{8}
\end{equation}%
This is due to the fact the upper limit in the sum $\sum%
\limits_{j=1}^{v_{0}+t-1}$ depends on $t$. This can be clarified further if,
in the continue limit, instead of the sum

\begin{equation}
K\equiv \sum \limits_{j=1}^{v_{0}+t-1}k_{j},  \tag{9}
\end{equation}%
one writes

\begin{equation}
K\rightarrow \mathcal{K}=\sum
\limits_{j=1}^{v_{0}}k_{j}+\int_{t_{i}}^{t-1}k(t)dt.  \tag{10}
\end{equation}

The probability that a vertex has connectivity $k_{i}$ smaller than $k$ can
be written as

\begin{equation}
P(k_{i}(t)<k)=P(t_{i}>\frac{e^{2}t}{k^{2}}).  \tag{11}
\end{equation}%
Combining (7) and (11) we obtain

\begin{equation}
P(t_{i}>\frac{e^{2}t}{k^{2}})=1-P(t_{i}\leq \frac{e^{2}t}{k^{2}})=1-\frac{%
e^{2}t}{k^{2}(v_{0}+t)}.  \tag{12}
\end{equation}%
Here, we have assumed that the probability density for $t_{i}$ is $%
P(t_{i})=1/(v_{0}+t)$. So, we get

\begin{equation}
P(k)=\frac{\partial P(k_{i}(t)<k)}{\partial k}=\alpha \frac{1}{k^{3}}, 
\tag{13}
\end{equation}%
where%
\begin{equation}
\alpha =\left( \frac{2e^{2}t}{v_{0}+t}\right) .  \tag{14}
\end{equation}%
Comparing (1) with (13) one sees that in this model the free-scaling
parameter $\gamma $ becomes $\gamma =3$.

\smallskip \ 

\noindent \textbf{3. Gravitational information theory}

\smallskip \ 

Recently, in Ref. [10] it has been shown that Newton's law of gravity can be
obtained from information theory. The central idea is to assume that the
space, in which one considers the motion of particles of mass $m$, is a
storage of information and that this information can be storage in certain
surfaces or screens. In particular one may assume that such a surface
corresponds to a sphere $S^{2}$. Moreover, the information is measure by
bits. Thus, one assumes that the number of bits $N$ storage in a sphere is
proportional to the area $A$, that is

\begin{equation}
N=\frac{A}{l_{p}^{2}},  \tag{15}
\end{equation}%
where%
\begin{equation}
A=4\pi r^{2},  \tag{16}
\end{equation}%
and

\begin{equation}
l_{p}=\sqrt{\frac{G\hslash }{c^{3}},}  \tag{17}
\end{equation}%
are the area of a sphere and the Planck's length, respectively.

Thus using the thermodynamic relation between the force $F$ and the
temperature $T$,

\begin{equation}
F=(\frac{2\pi k_{B}mc}{\hslash })T,  \tag{18}
\end{equation}%
the equipartition rule for the energy

\begin{equation}
E=\frac{1}{2}Nk_{B}T,  \tag{19}
\end{equation}%
and the rest mass equation

\begin{equation}
E=Mc^{2},  \tag{20}
\end{equation}%
one obtains that

\begin{equation}
F=G\frac{Mm}{r^{2}},  \tag{21}
\end{equation}%
which is the familiar Newton's law of gravitation. Here, $M$ denotes the
mass enclosed by a spherical screen $S^{2}$ (see Ref. [10] for details).

\smallskip \ 

\noindent \textbf{4. Gravitational complex network}

\smallskip \ 

We shall now combine the results of the section 2 and 3. The central idea is
to link (13) and (21). For this purpose let us write (13) and (21) in form

\begin{equation}
P\sim \frac{1}{k^{3}},  \tag{22}
\end{equation}%
and%
\begin{equation}
F\sim \frac{1}{r^{2}},  \tag{23}
\end{equation}%
respectively. It is evident that these expressions suggest the
identifications $P\longleftrightarrow F$. Consequently one discovers the
possible relation 
\begin{equation}
r\sim k^{3/2},  \tag{24}
\end{equation}%
between the radio $r$ and the connectivity $k$.

However the expression (22) is just one of many possibilities [11]. In
general, one should have

\begin{equation}
P\sim \frac{1}{k^{\gamma }},  \tag{25}
\end{equation}%
where, as it was mentioned in section 1, $\gamma $ is just a free-scale
parameter called the connectivity distribution exponent.

It turns out that the scale-free parameter $\gamma $ is a model dependent.
For instance changing the preferential axiom mentioned in section 2, $\gamma 
$ can have values between $2$ and infinity. However, in the observed
networks the values of $\gamma $ fall only between $2$ and $3$. An
interesting possibility to explain this phenomena was proposed in Ref. [12].
According to this work few scale-free networks are observed because there
exists a natural boundary (cut-off) for the observation of the scale-free
networks.

For our case perhaps the most interesting case is when $\gamma =2$, because
(25) becomes%
\begin{equation}
P\sim \frac{1}{k^{2}},  \tag{26}
\end{equation}%
and therefore one can make the identification

\begin{equation}
r\sim k,  \tag{27}
\end{equation}%
which is simpler than (24).

We would like to emphasize, the important role played by formula (15) in
these connections. This is a key formula because it allows us to consider
the parameter $r$ as a discrete statistic quantity. In fact, thanks to this
formula one may identify a random $r$ with a random connectivity $k$ as in
(27).

\smallskip \ 

\noindent \textbf{5. Some general comments}

\smallskip

In the previous section it was assumed that the connectivity $k$ is a
continuous real variable. But one may wonder whether there exist models that
it do not use the continuum assumption. In fact, there are two equivalent
approaches, namely the master-equation [13] and the rate-equation approach
[14]. In the first case one considers the probability $p(k,t_{i},t)$ that at
time $t$ a node $i$, introduced a time $t_{i}$, has a degree $k$. The master
equation is

\begin{equation}
p(k,t_{i},t+1)=\frac{k-1}{2t}p(k-1,t_{i},t)+(1-\frac{k}{2t})p(k,t_{i},t). 
\tag{28}
\end{equation}%
It turns out that the degree distribution $P(k)$ can be obtained from $%
p(k,t_{i},t)$ through the formula

\begin{equation}
P(k)=\lim_{t\longrightarrow \infty }\frac{(\sum_{t_{i}}p(k,t_{i},t))}{t}, 
\tag{29}
\end{equation}%
(see Ref. [13] for details). In the second case, one focuses on the average $%
N_{k}(t)$ of nodes with $k$ edges a time $t$. The rate equation for $N_{k}(t)
$ is

\begin{equation}
\frac{dN_{k}(t)}{dt}=m\frac{(k-1)N_{k-1}(t)-kN_{k}(t)}{\sum_{k}kN_{k}(t)}%
+\delta _{km}.  \tag{30}
\end{equation}%
In the asymptotic limit one has

\begin{equation}
N_{k}(t)=tP(k),  \tag{31}
\end{equation}%
(see Ref. [14] for details). What it is important is that these two
approaches are equivalent and that both lead to the continuum theory in the
asymptotic limit.

The identification of $k\sim r$ given in (27) deserves additional comments.
The connectivity $k$ refers to the number of edges in a given vertex of a
graph $G$. So if we may relate $r$ with a given graph $G$ we will be closed
to clarifies such a connection. Following Verlinde [10] let us assume that
the screen associated to the mass $m$ is a sphere $S^{2}$. This sphere has
radius $r$ and area $A=4\pi r^{2}$. The central idea in emergent gravity is
to visualize such a sphere $S^{2}$ as storage of information in the form of $%
N$ bits, which are linked to $A$ according to the formula (15). From
topology, we know that a sphere $S^{2}$ is triangulable. This means that a
sphere is homeomorphic to the corresponding polyhedron. It turn out that by
a stereographic projection one knows that $S^{2}\sim R^{2}\cup \{ \infty \}$.
This means that one can visualize the polyhedron associated with $S^{2}$ as
a connected graph $\mathcal{G}$ drawing in the plane $R^{2}\cup \{ \infty \}$%
. The equator of $S^{2}$ is a circle $S^{1}$ with radius $r$. So our task is
to see whether $r$ can be related to a kind of distance $d(v_{i},v_{j})$
connecting to vertices $v_{i}$ and $v_{j}$ of the given graph $\mathcal{G}$
in the plane. Fortunately, in Ref. [15] it is discussed an information
processing in complex networks precisely by introducing the shortest
distance $d(v_{i},v_{j})$ between vertices $v_{i}$ and $v_{j}$. Moreover, in
a such reference the $j$-sphere is defined as

\begin{equation}
S_{j}(v_{i},\mathcal{G})=\{v\mid d(v,v_{j})=j,\text{ }j\geq 1\}.  \tag{32}
\end{equation}%
By defining the information functional of a graph $\mathcal{G}_{P}$ , $f:%
\mathcal{G}_{P}\longrightarrow R_{+}$, the vertex probability

\begin{equation}
p(v_{i})=\frac{f(v_{i})}{\sum_{i}f(v_{i})}  \tag{33}
\end{equation}%
can be introduced. Here $\mathcal{G}_{P}$ is constructed from the paths $%
P_{k_{j}}^{j}(v_{i})$ and the associated edges $E_{k_{j}}$ sets of the set

\begin{equation}
S_{j}(v_{i},\mathcal{G})=\{v_{u_{j}},v_{v_{j}},...,v_{x_{j}}\}.  \tag{34}
\end{equation}%
It turns out that the functional $f$ captures structural information of the
underlaying graph $\mathcal{G}$ (See Ref. [15] for details.) Going backwards
it must be possible to prove that such a structural information of the graph 
$\mathcal{G}$ in the plane $R^{2}\cup \{ \infty \}$ must be linked to the the
bits $N$ storage on the sphere $S^{2}$.

\smallskip \ 

\noindent \textbf{6. Final remarks}

\smallskip

Our proposed bridge between growing networks and gravity may help to develop
the corresponding formalism in both directions. For instance, starting with
growing networks and using (27) or (24) one may be able to rediscover the
thermodynamic view of gravity. On the other hand starting with gravity one
may bring concepts, such as geometry, in to the scenario of evolving
networks. And in this direction, perhaps one may be able to speak of black
holes in growing networks. It is tempting to speculate that one may even
have a kind of Schwarzschild metric for complex networks of the form

\begin{equation}
ds^{2}=-(1-\frac{\beta }{k})dt^{2}+\frac{dk^{2}}{(1-\frac{\beta }{k})}%
+k^{2}(d\theta ^{2}+sen^{2}\theta d\phi ^{2}).  \tag{35}
\end{equation}%
Of course, in the context of complex networks one can raise many interesting
questions from this proposal, but honestly we do not have any idea what
could be the answer of such a questions. For instance, thinking about the
World Wide Web network of internet, what is it meaning of the concept of a
black hole? and in particular, what is it the meaning of the corresponding
event horizon associated with (28)? These are topics of great interest that
we leave for further research.

There are also a number of attractive directions where our work may find
some interest. In particular it may appear interesting to relate our work
with matroid theory [16] (see also Refs. [17]-[18] and references therein).
This is because graphs can be understood as a particular case of matroids
[19]-[20] and because in this case the concept of duality plays a
fundamental role. So one wonders if matroid-complex networks fusion (see
Ref. [21]) may bring eventually interesting and surprising results in
quantum gravity [22].

\  \bigskip

\noindent \textbf{Acknowledgments: }I would like to thank M. C. Mar\'{\i}n
and A. Le\'{o}n for helpful comments and the Mathematical, Computational \&
Modeling Sciences Center of the Arizona State University for the hospitality.

\smallskip \

\end{document}